\documentclass[amssymb,prb,aps,twocolumn,amsmath,showpacs]{revtex4}

\usepackage[dvips]{graphicx}
\usepackage{amstext}

\begin{document}

\title{Proximity-induced density-of-states oscillations in a superconductor/strong-ferromagnet system}
\author{Kurt M. Boden, W.P. Pratt, Jr. and Norman O. Birge}
\email{birge@pa.msu.edu}
\affiliation{Department of Physics and
Astronomy, Michigan State University, East Lansing, Michigan
48824-2320, USA}
\date{\today}

\begin{abstract}

We have measured the evolution of the tunneling density of states
(DOS) in superconductor/ferromagnet (S/F) bilayers with increasing
F-layer thickness, where F in our experiment is the strong
ferromagnet Ni.  As a function of increasing Ni thickness, we
detect multiple oscillations in the DOS at the Fermi energy from
differential conductance measurements.  The features in the DOS
associated with the proximity effect change from normal to
inverted twice as the Ni thickness increases from 1 to 5 nm.

\end{abstract}

 \maketitle

Hybrid systems consisting of superconducting (S) and ferromagnetic
(F) materials have attracted substantial attention due to their
interesting properties and potential for
applications.\cite{BuzdinReview}  The superconducting proximity
effect in such systems is normally short-ranged, due to the large
exchange energy in the F material. When a Cooper pair crosses the
S/F interface, the spin-up and spin-down electrons enter into
different spin bands, and the center of mass coordinate picks up
an oscillatory factor.\cite{Demler} The physical manifestations of
this oscillation can be observed as a series of transitions
between ``0" and ``$\pi$" states in S/F/S Josephson junctions as a
function of increasing F-layer
thickness,\cite{Kontos:02,Oboznov:06} or as oscillations between
``normal" and ``inverted" proximity features in the tunneling
density of states (DOS) of S/F/I/N tunnel
junctions.\cite{Kontos:01} (Here I is an insulator and N is a
normal metal.)

With substantial experimental effort in S/F/S Josephson junctions,
the 0-$\pi$ transition has been confirmed by many experimental
groups.\cite{Kontos:02, Oboznov:06, Ryazanov:01, Blum:02,
Sellier:03, Shelukhin:06, Weides:06, Robinson:06, Khaire:09}  What
is surprising is that, unlike in the Josephson geometry, the
oscillatory behavior of the DOS in S/F/I/N structures has been
observed convincingly only once, as a single normal-inverted
transition in samples with a weakly-ferromagnetic alloy for
F.\cite{Kontos:01} In experiments using strong ferromagnets, the
results have been less clear.\cite{Reymond:2006, SanGiorgio:2008}
At this time, to the best of our knowledge, there is no definitive
experimental answer to the question of whether the DOS in S/F/I/N
structures oscillates as a function of F-layer thickness when F is
a strong ferromagnet. The primary goal of this Rapid Communication
is to answer this question.
\begin{figure}[tbh]
\begin{center}
\includegraphics[width=3.0in]{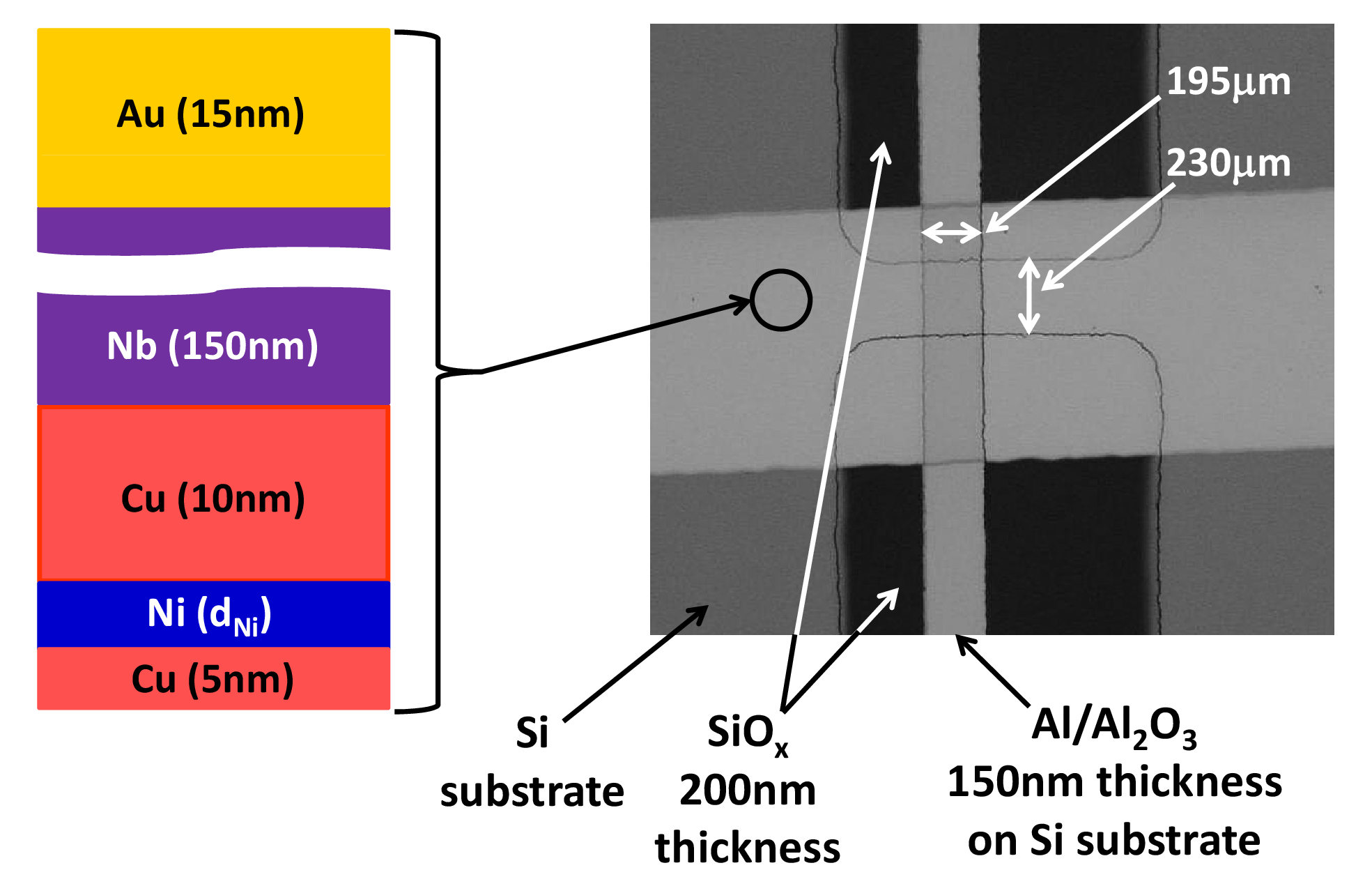}
\end{center}
\caption{(Color online).  (Left-hand side) Vertical representation
of the top lead (thin film multilayer) with associated
thicknesses. The functions of the various layers are described in
the text. The only variable thickness is Ni, d$_{Ni}$. (Right-hand
side) Top view of S/F/I/N junction using an optical microscope.
The junction area $A$ is defined by the edges of the SiO$_x$ and
the Al/Al$_2$O$_3$ bottom lead. The vertical and horizontal arrows
illustrate the junction width and length.}\label{Fig_1}
\end{figure}

Our S/F/I/N tunnel junctions are fabricated by thermal evaporation
and sputtering, using a series of mechanical masks (Fig. 1). We
first evaporate a 150 nm strip of Al (N), then we immediately
backfill the chamber with 300 Torr of a 10\% O$_2$, 90\% Ar
mixture. Exposing the freshly evaporated Al to the O$_2$ quickly
(while the Al is still hot) provides good conditions for oxide
growth. The O$_2$ exposure continues for $\approx$ 12 h to produce
a robust layer of Al$_2$O$_3$ (I) on the Al surface. Next, we
change masks and evaporate a thick layer (200 nm) of SiO$_x$ to
define the junction geometry. When using mechanical masks for the
top leads, shadow effects can cause unwanted regions at the edges
of the junctions where the Ni thickness is not well defined. The
SiO$_x$ is in place to avoid the appearance of edge effects in our
data. Finally, we sputter a
Cu(5nm)/Ni($x$)/Cu(10nm)/Nb(150nm)/Au(15nm) multilayer. The choice
of Ni and Cu is beneficial because our Ni has a relatively long
spin-diffusion length, $21 \pm 2$ nm, as compared to our maximum
Ni thickness, along with a low resistivity $\rho_{Ni} = 33 \pm 3$
n$\Omega$m.\cite{Moreau:2007}  Ni also provides weak asymmetry of
spin-dependent scattering in the bulk and at Ni/Cu interfaces, and
a low average Ni/Cu interface resistance.\cite{Moreau:2007} These
attributes should simplify theoretical analysis.  The Cu layer
adjacent to the Al$_2$O$_3$ has been found to increase the
effectiveness of the tunnel barrier. The Au deters oxidation of
the Nb layer. Throughout the process we must break vacuum, but the
consistency and reproducibility of our results suggest that this
has little effect on the quality of our junctions.  Due to a high
level of oxygenation of our Al, its T$_c \approx 1.9$ K. Thus, we
performed our measurements at 2.1 K, with the Al in the normal
state.

Using a four-terminal lock-in technique, we measure the
voltage-dependent differential conductance of our samples,
$dI/dV(V)$, which approximates the DOS of our S/F bilayer. (The
true DOS would be attained if measured at $T=0$ K when there is no
zero-bias anomaly; we will use ``DOS" to refer to our non-ideal
differential conductance measurements.) We normalize $dI/dV$ by
multiplying by the normal-state resistance $R_N$, determined from
the inverse of the differential conductance at large bias
voltages. Our junction area-resistance products, $AR_N \approx
2\times10^5$ $\Omega \mu$m$^2$, are substantially higher than those
of other groups using passive oxidation,\cite{Otto:2007,
Pothier:1997} but much lower than those produced by specialized
oxidation techniques.\cite{Kontos:01, SanGiorgio:2008} From the
data, it can be seen that this has little to no effect on our
measurements.
\begin{figure}[tbh]
\begin{center}
\includegraphics[width=3.2in]{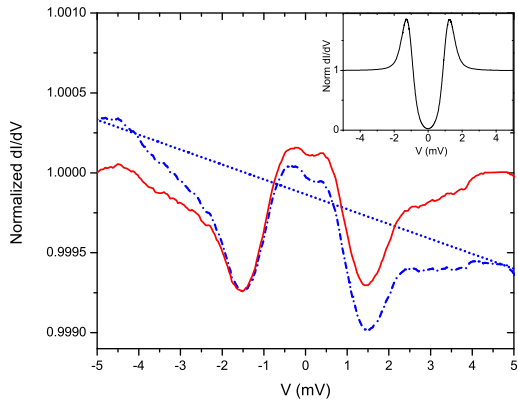}
\end{center}
\caption{(Color online).  The blue dot-dashed curve shows $dI/dV$
vs. $V$ for a d$_{Ni}$ = 4.5 nm junction, where the data are
slightly smoothed and normalized. The blue dotted line is a linear
fit to large bias voltage for the previous blue dot-dashed curve.
The red solid curve shows the same $dI/dV$ data after subtracting
off the linear background feature.  Inset: Normalized $dI/dV$ vs.
$V$ for an N/I/S junction with a 15-nm Cu buffer layer between S
and I. All data were taken at 2.1 K.}\label{Fig_2}
\end{figure}

Our $dI/dV$ data exhibit a small negatively sloped, linear
background, which appears to be a component of the normal state of
our junctions. Figure 2 shows a plot of $dI/dV$ from our 4.5-nm
sample before and after subtraction of the linear background,
along with the linear fit to $dI/dV$ for $|V| \gg \Delta$, where
$\Delta \approx 1.4$ mV is the gap parameter for Nb. There is
another normal state characteristic which we do not correct for in
our measurements.  A slight V-shaped feature centered at 0 V
becomes apparent at large Ni thicknesses. (It is visible on the
red solid curve of Fig. 2 at $|V|
> 2.5$ mV.) We did not have a magnet on our apparatus to force the top Nb layer
into the normal state; nevertheless, we emphasize that the slight
negative slope and V-shaped feature in our background are both
much smaller than background features observed in S/F/I/N tunnel
junctions measured by other groups.\cite{KontosThesis,
SanGiorgioThesis}

The inset of Fig. 2 shows a plot of $dI/dV$ for an N/I/S junction
with a 15-nm Cu buffer layer between the Nb and Al$_2$O$_3$
layers. The standard N/I/S junction behavior illustrates the
quality of the insulating barrier. The Cu buffer layer thickness
in the N/I/S sample is the sum of the Cu layer thicknesses in the
S/F/I/N samples. This was chosen to illustrate the minor effect of
the Cu in the S/F/I/N samples as a whole.  Even with this rather
thick layer of Cu and the elevated temperature, we see a gap where
$dI/dV$ nearly goes to zero. Another sign of junction quality is
that $dI/dV$ is featureless for $|V| \gg \Delta$ up to 10 mV (not
shown), the maximum measured $|V|$.
\begin{figure}[tbh]
\begin{center}
\includegraphics[width=3.5in]{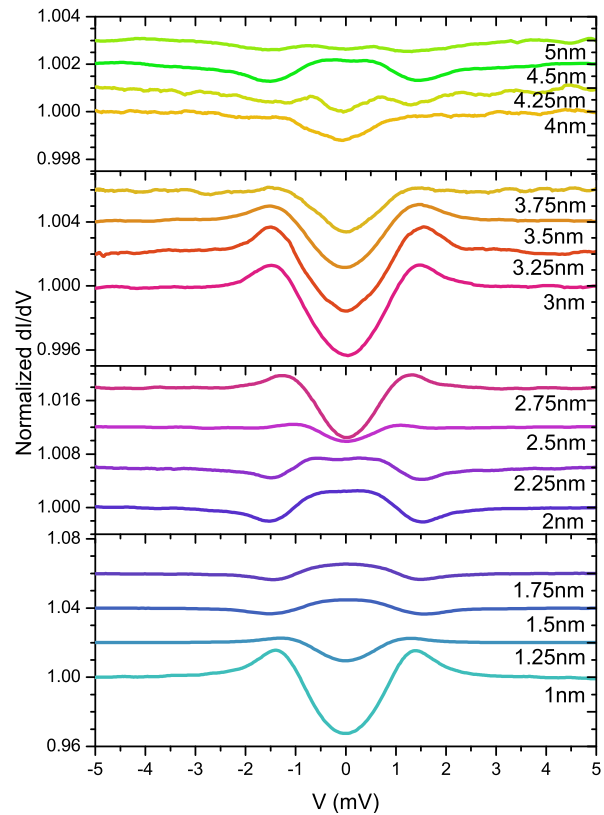}
\end{center}
\caption{(Color online).  Normalized differential conductance vs.
voltage for S/F/I/N junctions with several different Ni
thicknesses. The Ni thickness, $d_{Ni}$, is labeled under each
curve at the right. As one moves up the figure, each panel has an
increasing expansion of the vertical scale where only the lowest
trace is normalized to 1 and the others are displaced upward for
clarity. All data are taken at $T=2.1$ K.}\label{Fig_3}
\end{figure}

Figure 3 shows a plot of the DOS of our tunnel junctions while
varying the Ni thickness.  In a separate experiment at $T=5$ K, we
measured the saturation magnetization of Ni($d_{Ni}$)/Cu(5 nm)
multilayers, for 1 nm$<d_{Ni}<$5 nm.  The data show an
extrapolated non-magnetic ``dead-layer" Ni thickness of $0.25 \pm
0.05$ nm at each Ni/Cu interface.  Thus we show data only for
$d_{Ni}\geq 1$ nm. In the 1-nm sample one clearly sees the Nb gap,
but with a significant suppression of the bulk Nb features due to
the proximity effect in the strong ferromagnet. As we increase the
Ni thickness, the zero-bias dip in the DOS quickly decreases in
magnitude.  At $d_{Ni}=1.5$ nm, we observe the first sign of an
inversion in the differential conductance at zero bias, followed
by a maximum inversion at $d_{Ni}=1.75$ nm.  The features in
$dI/dV$ at $|V| = \Delta \approx 1.4$ mV have also been inverted
but occur at the Nb gap voltage in all the samples measured. The
inversion cycles quickly and by $d_{Ni}=2.5$ nm, the samples
return to the non-inverted regime. The dip in the DOS reaches its
maximum at $d_{Ni}=3$ nm, then a second inversion occurs starting
at $d_{Ni}\approx 4.25$ nm. (The second inversion is more apparent
in the expanded scale of Fig 2.) This second inverted state looks
as though it might extend past 5 nm.
\begin{figure}[tbh!]
\begin{center}
\includegraphics[width=3.2in]{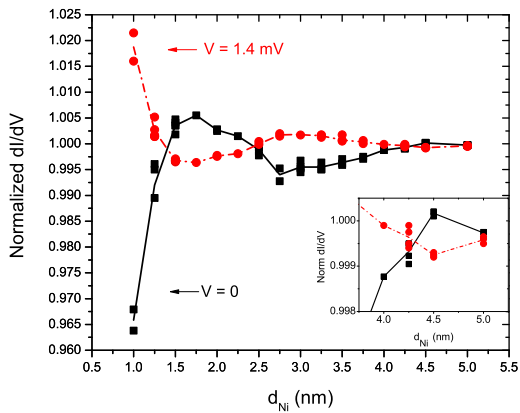}
\end{center}
\caption{(Color online).  Black squares represent the normalized
$dI/dV$ at $V=0$ for several Ni thicknesses.  Red circles
represent the normalized $dI/dV$ at $V=1.4$ mV. The black and red
lines link the average values of each thickness for $V=0$ and
$V=1.4$ mV, respectively. Inset: Data for $d_{Ni} \geq 4$ nm on an
expanded vertical scale.}\label{Fig 4}
\end{figure}

The oscillation is best illustrated in Fig 4 with a plot of the
normalized differential conductance at both $V=0$ and $V=1.4$ mV
versus Ni thickness.  The oscillation period is irregular; the
first inverted region persists for $d_{Ni} \approx 1.3 - 2.4$ nm,
while the second non-inverted region lasts much longer, for
$d_{Ni} \approx 2.4 - 4.3$ nm. One can see in the inset that the
oscillation occurs about a value $dI/dV \approx 0.9996$. This
offset in the oscillation is due to the aforementioned, weak
V-shaped zero-bias anomaly.

The transitions between normal and inverted DOS have been
predicted to occur at F-layer thicknesses exactly half of where
0-$\pi$ transitions occur in S/F/S Josephson
junctions.\cite{BuzdinReview}  After reviewing the literature on
S/Ni/S junctions, we see that our junctions are most similar to
those of Blum \textit{et al.}\cite{Blum:02} and Shelukhin
\textit{et al.},\cite{Shelukhin:06} where a Cu buffer layer is
placed on each side of the Ni layer. In contrast, Robinson
\textit{et al.}\cite{Robinson:06} have Nb in direct contact with
Ni, with a Ni ``dead-layer" thickness of $\approx$ 0.8 nm at each
Ni/Nb interface in comparison to our Ni dead-layer thickness of
only $0.25 \pm 0.05$ nm at each Cu/Ni interface.   The Ni
thickness at which the first 0-$\pi$ transition is observed by
these groups varies quite a bit -- 2.6 nm, 1.7 nm (extrapolated
value), and 3.8 nm, for Refs. [\onlinecite{Blum:02}],
[\onlinecite{Shelukhin:06}], and [\onlinecite{Robinson:06}],
respectively. Since our first 0-$\pi$ and $\pi$-0 transitions
occur at 1.3 and 2.5 nm, respectively, we would expect the first
two transitions in S/Ni/S junctions to occur at 2.6 and 5 nm, in
reasonable agreement with the values observed by Blum \textit{et
al.}\cite{Blum:02}

Theoretical calculations of the DOS in S/F bilayers cover several
regimes, defined by the relative strengths of the ferromagnetic
exchange, $E_{ex}$, the superconducting gap parameter, $\Delta$,
and the impurity scattering, $\hbar/\tau_e$, as well as the
relative sizes of $d_F$ and the mean free path, $l_e = v_F
\tau_e$. In the dirty limit, the Usadel equations provide a
clearcut prediction of oscillation of the tunneling DOS, with
period of order $\xi_F = (\hbar D/E_{ex})^{1/2}$, where $D=v_F
l_e/3$.\cite{Buzdin:2000} In the clean limit, the predictions are
less straightforward. Solving the Eilenberger equation in the
ballistic limit leads to the conclusion that the DOS does not
oscillate in a semi-infinite ferromagnet.\cite{Buzdin:2001,
Bergeret:2002} Oscillations are predicted to occur, however, in
the presence of weak disorder, with an amplitude proportional to
$\hbar/(E_{ex}\tau_e)$.\cite{Buzdin:2001, Bergeret:2002} With a
finite ferromagnet, the oscillation is regained even in the pure
ballistic limit due to specular\cite{Halterman:2002} or
diffuse\cite{Zareyan:2001} scattering from the film boundaries. In
constrast, Sun \textit{et al.} solved the Bogoliubov-deGennes
equations and claimed that oscillations should occur in either a
finite or semi-infinite F layer.\cite{Sun:2002}

Because our observed DOS variations are not periodic in $d_{Ni}$,
we do not attempt to fit our data with Usadel theory. We believe
that our samples most closely match the assumptions in the papers
by Zareyan \textit{et al.}\cite{Zareyan:2001} From earlier work on
S/Ni systems, we expect that $E_{ex} \approx 100$
meV;\cite{Blum:02, Shelukhin:06, Robinson:06} and $v_F = 2.8
\times 10^5$ m/s for Ni,\cite{Petrovykh:98} while $\Delta = 1.5$
meV for Nb. From our measured Ni resistivity, $\rho_{Ni} = 33$
n$\Omega$m, we deduce $l_e^{Ni} = 45$ nm. This puts our samples in
the ``intermediate" regime with $E_{ex} \gg \hbar/\tau_e \gg
\Delta$. By using Cu buffer layers next to the Ni, we limit the
scattering events in our junctions considerably. As stated
earlier, we find a spin diffusion length of 21 $\pm 2$ nm, low
overall spin-scattering asymmetry, and very low Cu/Ni interface
specific resistance in our multilayers: $AR_{Cu/Ni} = 0.18 \pm
0.03 f\Omega m^2$.\cite{Moreau:2007} This low interface resistance
corresponds to a probability of scattering of only $\sim$15\% at
each Ni/Cu interface.\cite{ScatteringCalc} The Cu/Nb interface is
``rough" in the sense that there is significant diffusive
scattering at this interface, as determined from its measured
interface specific resistance of $AR_{Cu/Nb} = 1.1 \pm 0.15
f\Omega m^2$.\cite{Park:2000} This value of $AR_{Cu/Nb}$ is larger
than the total $AR = 2AR_{Cu/Ni} + \rho_{Ni}d_{Ni} +
\rho_{Cu}d_{Cu}$ = 0.54 and 0.67 $f\Omega m^2$ of the Cu/Ni/Cu
region for $d_{Ni}$ = 1 and 5 nm, respectively. We also expect
there to be diffuse scattering at the Cu/Al$_2$O$_3$ (tunneling)
interface.

Plots of the energy dependence of the DOS shown in the papers by
Zareyan \textit{et al.}\cite{Zareyan:2001} agree qualitatively
with our data. Performing a quantitative fit of the theory to our
data, however, is problematic. The theory predicts that the first
0-$\pi$ transition should occur at very small $d_{Ni}$, a flaw
that may be correctable by adding spin-dependent interfacial phase
shifts to the theory.\cite{Cottet:2005} (One could also argue
that, because of the $0.25$-nm dead layers at the two Cu/Ni
interfaces, one should subtract 0.5 nm from our nominal sample
thickness before fitting to the Zareyan theory, but that is not
nearly enough to bring theory into agreement with experiment.) The
theory also predicts large oscillations in the normalized DOS
(i.e., large deviations from 1) at zero energy -- much larger than
what we observe in the experiment. The amplitude of the
theoretical oscillations can be reduced by assuming a very small
transparency $T$ of the Nb/Cu interface; such an assumption,
however, is incompatible with the measured boundary resistance
$AR_{Cu/Nb} = 1.1 \pm 0.15 f\Omega m^2$, which implies that $T
\approx 0.5$.\cite{ScatteringCalc} Strong spin-flip scattering
would also reduce the amplitude of the DOS variations; the long
measured spin memory length in our Ni films, however, precludes
that explanation for these samples.  One could assume that the
variation in F-layer thickness over the junction area is very
large, thereby smearing out the oscillations; we believe that such
an assumption is unrealistic.

A previous measurement of the Nb/Ni system\cite{SanGiorgio:2008}
did detect signs of one $0-\pi$ transition, but the data contained
additional low-energy features, which were later interpreted as
signs of p-wave spin-triplet pairs.\cite{Lu:2010}  We do not
observe such low-energy features in our data.

In conclusion, we have observed multiple oscillations in the DOS
of S/F bilayers as a function of F-layer thickness, where F is a
strong ferromagnet (Ni).  The oscillations can be described
qualitatively, but not quantitatively, by the theory of Zareyan
\textit{et al.}.\cite{Zareyan:2001}  Discrepancies between theory
and experiment may be due to the extra Cu layers in our samples,
which are not present in the theoretical calculation, or to the
absence of spin-dependent interfacial phase shifts in the
theory.\cite{Cottet:2005}

We acknowledge helpful conversations with M. Aprili and W. Belzig,
technical assistance from R. Loloee and B. Bi, and use of the W.M.
Keck Microfabrication Facility. This work was supported by the
U.S. Department of Energy under Grant No. DE-FG02-06ER46341.


\begin{thebibliography} {99}

\bibitem{BuzdinReview} A.I. Buzdin, Rev. Mod. Phys. {\textbf{77}} 935 (2005).

\bibitem{Demler} E.A. Demler, G.B. Arnold, and M.R. Beasley,
Phys. Rev. B \textbf{55}, 15174 (1997).

\bibitem{Kontos:02} T. Kontos, M. Aprili, J. Lesueur, F. Genet,
B. Stephanidis, and R. Boursier, Phys. Rev. Lett. \textbf{89}, 137007 (2002).

\bibitem{Oboznov:06} V.A. Oboznov, V.V. Bol'ginov, A.K.
Feofanov, V.V. Ryazanov, and A.I. Buzdin, Phys. Rev. Lett.
\textbf{96}, 197003 (2006).

\bibitem{Kontos:01} T. Kontos, M. Aprili, J. Lesueur, and X.
Grison, Phys. Rev. Lett. \textbf{86}, 304 (2001).

\bibitem{Ryazanov:01} V.V. Ryazanov, V.A. Oboznov, A.Yu. Rusanov, A.V. Veretennikov, A.A. Golubov and
J. Aarts, Phys. Rev. Lett. \textbf{86}, 2427 (2001).

\bibitem{Blum:02} Y. Blum, A. Tsukernik, M. Karpovski, and A.
Palevski, Phys. Rev. Lett. \textbf{89}, 187004 (2002).

\bibitem{Sellier:03} H. Sellier, C. Baraduc, F. Lefloch, and R.
Calemczuk, Phys. Rev. B \textbf{68}, 054531 (2003).

\bibitem{Shelukhin:06} V. Shelukhin, A. Tsukernik, M. Karpovski,
Y. Blum, K.B. Efetov, A.F. Volkov, T. Champel, M. Eschrig, T.
L$\ddot{o}$fwander, G. Sch$\ddot{o}$n, and A. Palevski, Phys. Rev.
B \textbf{73}, 174506 (2006).

\bibitem{Weides:06} M. Weides, M. Kemmler, E. Goldobin, D. Koelle,
R. Kleiner, H. Kohlshedt, and A. Buzdin, Appl. Phys. Lett.
\textbf{89}, 122511 (2006).

\bibitem{Robinson:06} J.W.A. Robinson, S. Piano, G. Burnell, C.
Bell and M.G. Blamire, Phys. Rev. Lett. \textbf{97}, 177003
(2006); Phys. Rev. B \textbf{76}, 094522 (2007).

\bibitem{Khaire:09} T.S. Khaire, W.P. Pratt Jr. and N.O. Birge, Phys. Rev. B \textbf{79}
094523(2009).

\bibitem{Reymond:2006}S. Reymond, P. SanGiorgio, M.R. Beasley, J. Kim, T. Kim, and K. Char, Phys. Rev. B \textbf{73}, 054505 (2006).

\bibitem{SanGiorgio:2008} P. SanGiorgio, S. Reymond, M.R. Beasley,
J.H. Kwon, and K. Char, Phys. Rev. Lett. \textbf{100}, 237002
(2008).

\bibitem{Moreau:2007} C.E. Moreau, I.C. Moraru, N.O. Birge and W.P. Pratt, Jr.,
Appl. Phys. Letters \textbf{90}, 012101 (2007).

\bibitem{Otto:2007} E. Otto, M. Tarasov, and L. Kuzmin,
J. Vac. Sci. Technol. B, \textbf{25}, No. 4 (2007).

\bibitem{Pothier:1997} H. Pothier, S. Gueron, N.O. Birge, D. Esteve, and M.H. Devoret, Phys. Rev. Lett. \textbf{79}, 3490 (1997).

\bibitem{KontosThesis} T. Kontos, Ph.D. thesis, Universit\'e Paris XI, 2002.

\bibitem{SanGiorgioThesis} P. SanGiorgio, Ph.D. thesis, Stanford University, 2008.

\bibitem{Buzdin:2000} A. Buzdin, Phys. Rev. B \textbf{62}, 11377 (2000).

\bibitem{Buzdin:2001} I. Baladie, A. Buzdin, Phys. Rev. B \textbf{64}, 224514 (2001).

\bibitem{Bergeret:2002} F. S. Bergeret, A. F. Volkov, and K. B. Efetov, Phys. Rev. B \textbf{65}, 134505 (2002).

\bibitem{Halterman:2002} K. Halterman and O.T. Valls, Phys. Rev. B \textbf{66}, 224516 (2002).

\bibitem{Zareyan:2001} M. Zareyan, W. Belzig, and Yu. V. Nazarov, Phys. Rev. Lett. \textbf{86}, 308 (2001); Phys. Rev. B. \textbf{65}, 184505 (2002).

\bibitem{Sun:2002} G. Sun, D.Y. Xing, J. Dong, and M. Liu, Phys. Rev. B \textbf{65}, 174508 (2002).

\bibitem{Petrovykh:98} D.Y. Petrovykh, K.N. Altmann, H. Hochst, M.
Laubscher, S. Maat, G.J. Mankey, and F.J. Himpsel, Appl. Phys.
Lett. \textbf{73}, 3459 (1998).

\bibitem{ScatteringCalc} One can estimate the average transmission probability, $T$, at an
interface from the measured specific resistance, $AR$, using the
Landauer approach.  From Eqn. (6) in Schep \textit{et
al.}\cite{Schep:1997} one obtains the result $1/T - 1 =
AR*(2e^2/h)*(k_F^2/4\pi) \approx AR/(1 f\Omega m^2)$. For Cu/Ni,
this calculation gives $T \approx 0.85$, while for Nb/Cu it gives
$T \approx 0.5$.

\bibitem{Schep:1997} K.M. Schep, J.B.A.N. van Hoof, P.J. Kelly,
G.E.W. Bauer, and J.E. Inglesfield, Phys. Rev. B \textbf{56},
10805 (1997).

\bibitem{Park:2000} W. Park, D.V. Baxter, S. Steenwyk, I. Moraru, W. P. Pratt, Jr., and J. Bass, Phys. Rev. B \textbf{62}, 1178 (2000).

\bibitem{Cottet:2005} A. Cottet and W. Belzig, Phys. Rev. B \textbf{72}, 180503(R)
(2005); A. Cottet, \textit{ibid.} \textbf{76}, 224505 (2007).

\bibitem{Lu:2010} W.J. Lu, Y.K. Bang, and K. Char, Phys. Rev. B \textbf{81}, 144514 (2010).

\end{thebibliography}
\end{document}